\documentclass[aps,twocolumn,floatfix,groupedaddress]{revtex4-2}

\usepackage{graphicx}
\usepackage{dcolumn}
\usepackage{bm}
\usepackage{color} 
\usepackage{amssymb,amsmath} 

\newcommand{\T}[1]{\mathbf{$1$}}

\newcommand{\refeq}[1]{(\ref{#1})}

\newcommand{\ice}{{\mathit{ice}}}

\begin{document}

\preprint{LA-UR-22-2460}

\title{Influence of mass ablation on ignition and burn propagation in layered fusion capsules}


\author{W. Daughton}
\email[]{daughton@lanl.gov}
\author{B. J. Albright}
\author{S. M. Finnegan}
\author{Brian~M.~Haines}
\author{J. L. Kline}
\author{J. P. Sauppe}
\author{J. M. Smidt}
\affiliation{Los Alamos National Laboratory, Los Alamos, New Mexico 87545}

\date{\today}

\begin{abstract}
After decades of research, recent laser-driven inertial fusion experiments have demonstrated a rapid progress toward achieving thermonuclear ignition using capsule designs with cryogenic fuel layers.  The ignition physics for these layered capsules involves a complex interplay between the dynamically forming hot spot and the dense surrounding fuel.   Using analytic theory and numerical simulations, we demonstrate that the mass ablation rate into the hot spot depends sensitively upon the temperature of the dense fuel, resulting in ablative inflows up to $4\times$ faster than previous estimates.  This produces an enthalpy flux into the hot spot that plays a critical role in controlling the hot spot temperature, the ignition threshold, and the subsequent burn propagation. The net influence of mass ablation on the ignition threshold is regulated by a dimensionless parameter that depends upon the temperature of the dense fuel. As a consequence, the ignition threshold is sensitive to any mechanism that heats the dense fuel, such as neutrons or radiation emitted from the hot spot. These predictions are confirmed using radiation hydrodynamic simulations for a series of capsules near ignition conditions.  This analysis may have relevance for understanding the variable performance of recent experiments and for guiding new capsule designs toward higher fusion yields.
\end{abstract}


\maketitle

\section{Introduction}

A principal approach to inertial confinement fusion (ICF) employs lasers \cite{Nuckolls_1972} to compress and heat hydrogen isotopes to conditions where thermonuclear reactions produce self-heating of the plasma.  Recent experiments on the National Ignition Facility (NIF) have demonstrated hot spot ignition using indirect drive \cite{Hurricane_2022}.  This approach to ICF \cite{Lindl_1995}  employs 192 laser beams (1.91~MJ) directed into a gold (or uranium) hohlraum to generate x-ray radiation ($\sim300$ eV), which then induces rapid ablation on the outer surface of the capsule and drives high velocity ($\sim 400$ km/s) implosions.  The recent HYBRID-E capsules \cite{Hurricane_2022, Zylstra_2022, Zylstra_2021, Kritcher_2021, Ross_2022} employ high-density carbon (HDC) for the ablator with a shell of cryogenic deuterium and tritium (DT) ice layered on the inner surface.   At the cryogenic initial conditions, the central cavity contains a low density DT vapor which is a small fraction ($\sim 1\%$) of the total DT ice mass ($\sim 200 \mu\rm g$).   Such designs \cite{Lindl_1995} are attractive for achieving high-gain, by first igniting a central hot spot ($\rho \sim 100\; {\rm g/cm^{3}}$, ${\bar T}\sim 6$ keV) which can then propagate into the dense fuel.   However, the precise requirements for ignition and burn propagation are not fully understood.   Since the observed neutron yield \cite{Hurricane_2022}  corresponds to a small burn-up fraction ($\sim 2\%$) of the fuel, further design improvements are needed to achieve high-gain. 
  
One central feature of {\em layered capsules} is that the hot spot mass ($M \sim 40 \mu\rm g$) is formed dynamically from the ice during the implosion \cite{Guskov_1976,Atzeni_1984,Lindl_1995,Betti_2001}.  First, through shock and compressional heating, the DT vapor rapidly heats to $T \sim 3$~keV leading to rapid thermal conduction into the ice and driving mass ablation into the hot spot.  As the hot spot temperature and density increase,  the D+T fusion reactions produce 3.5~MeV $\alpha$-particles and 14.1~MeV neutrons.  Some fraction of the $\alpha$-energy is deposited directly into the hot spot, while the remainder is deposited in the surrounding dense fuel, driving additional mass ablation into the hot spot.    These ablative inflows recycle energy back into the hot spot, while thermal conduction and sound waves rapidly redistribute this energy, leading to an isobaric hot spot with a self-similar temperature profile \cite{Atzeni_1984,Betti_2001}.   Any new mass injected into the hot spot rapidly becomes thermally coupled and plays an important role in regulating the mass averaged temperature and the subsequent thermonuclear reactions.  While elements of this picture are widely accepted, the most commonly used expression for the mass ablation rate ($dM/dt$) \cite{Atzeni_1984, Atzeni_2004,Hurricane_2016,Hurricane_2019,Hurricane_2019B}  has not been systematically tested.  Several other models have focused on mass ablation driven by thermal conduction in the limit where the dense fuel is cold \cite{Betti_2001, Saillard_2006}.   However, it appears that all previous models \cite{Kidder_1976,Meyer-Ter-Vehn_1982,Atzeni_1984, Betti_2001,Atzeni_2004, Zhou_2008, Betti_2010,Cheng_2014,Springer_2019, Hurricane_2019,Hurricane_2019B, Christopherson_2020, Hurricane_2021,MacLaren_2021,Cheng_2021} have either assumed that the enthalpy flux is negligible or that it is balanced by the heat flux leaving the hot spot.  While this assumption may be partially justified when the dense fuel is cold, it is not appropriate near ignition conditions as the dense fuel is heated.   

Here, we demonstrate that the ablation rate into the hot spot depends sensitively upon the temperature of the dense surrounding fuel, which we denote as $T_\ice$ since this region originates from the ice layer.  While our analysis recovers the previous expression \cite{Atzeni_1984, Atzeni_2004,Hurricane_2016,Hurricane_2019,Hurricane_2019B} for $dM/dt$ in the limit $T_\ice \rightarrow 0$,  the predicted ablation rate is up to $4\times$ faster as the dense fuel is heated.   As a result, the ablative inflow produces an enthalpy flux into the hot spot that competes with the previously identified $dM/dt$ term \cite{Atzeni_1984,Betti_2001} in the power balance.  Accounting for both terms, the hot spot power balance is well satisfied in the simulations, which demonstrates the consistency of our analysis.   By combining $dM/dt$ terms, a dimensionless parameter is found that regulates the net influence of mass ablation on the ignition threshold near stagnation.  In addition to the hot spot temperature and the areal density, the new ignition threshold depends critically upon the temperature of the dense fuel, since this influences the ablation rate into the hot spot.  Near stagnation time, mass ablation plays an important role in limiting the hot spot temperature by increasing the heat capacity.   As a result,  the ignition threshold is significantly more pessimistic relative to previous estimates.  We demonstrate that these new predictions are consistent with radiation hydrodynamic simulations and with the experimentally observed ignition cliff in recent NIF capsules \cite{Hurricane_2022, Zylstra_2022}.  Thus, our results may have relevance for understanding why ignition occurred in the recent experiment \cite{Hurricane_2022} as well as the variable performance of repeat shots over the past year \cite{Tollefson_2022}.   

These results also have implications for the design of new experiments. While researchers have long recognized the importance of the dense fuel adiabat for achieving the required compression, our results demonstrate that the temperature of the dense fuel plays a key role in the power balance.  Any process that heats the dense fuel (radiation, neutrons) will increase the mass ablation rate into the hot spot and impact both the ignition threshold and subsequent burn propagation.   These conclusions may also be relevant to other types of layered capsule designs \cite{MacLaren_2021} including concepts with liquid DT layers in both spherical \cite{Olson_2016,Paddock_2020, Norreys_2020,Olson_2021} and cylindrical implosions \cite{slutz_2022}.

To improve the readability of this paper, an overview of the new ignition theory is presented in Sec.~\ref{theory}, while the detailed derivation of the mass ablation rate is given in Appendix~\ref{MassAppendix}, and the scalings employed for the ignition threshold are given in Appendix~\ref{IgnitionAppendix}.   The theoretical predictions are tested with radiation-hydrodynamics simulations in Sec.~\ref{radhydro} and conclusions are summarized in Sec.~\ref{summary}.

\section{Theoretical Synopsis}
\label{theory}
\begin{figure}
 \includegraphics[width=0.75\linewidth]{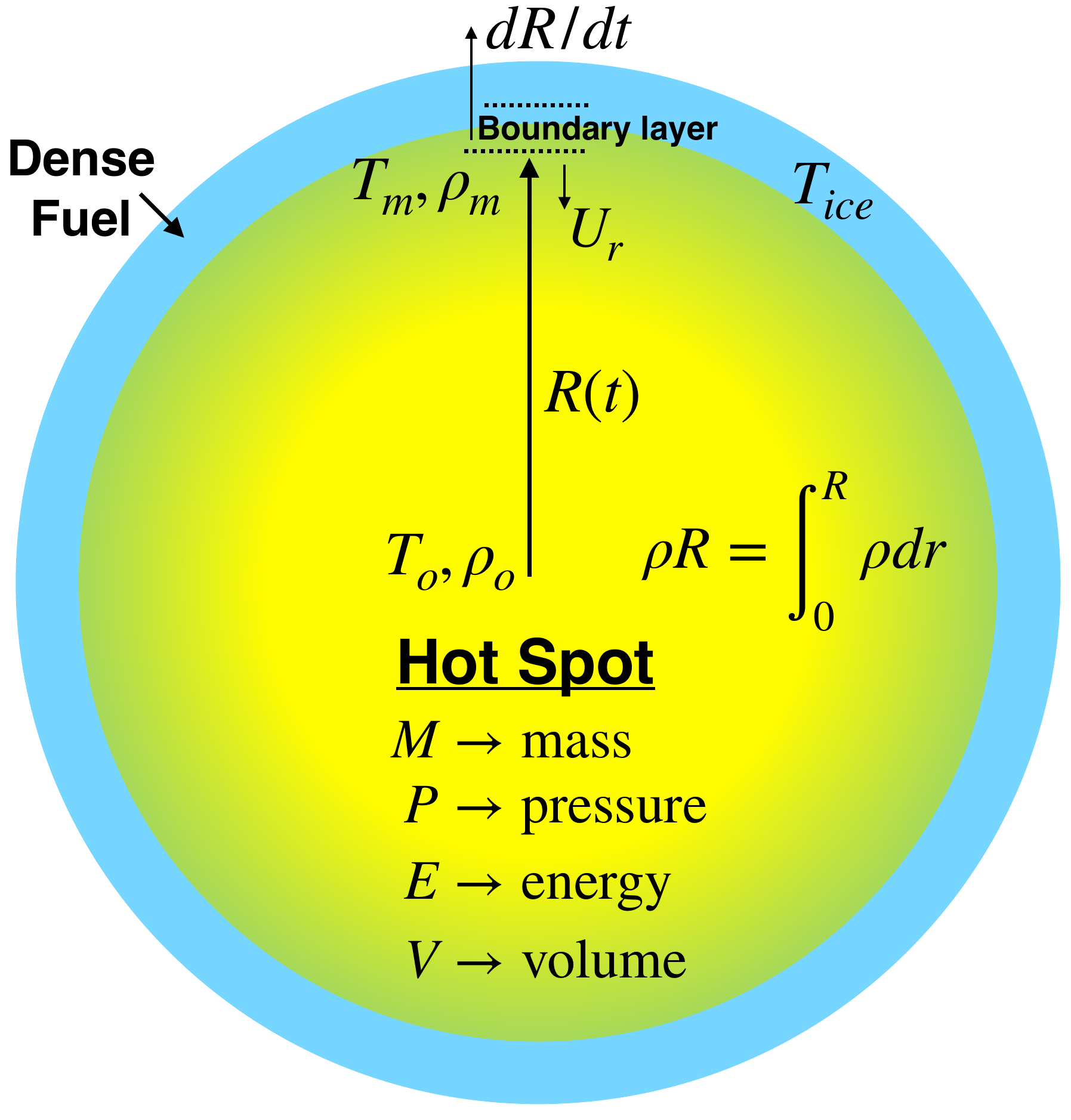}
 \caption{Notational sketch showing the density and temperature at the center ($\rho_o, T_o$) and at the matching location ($\rho_m, T_m$) corresponding to the radius $R(t)$.   The analysis in Appendix~\ref{MassAppendix} matches the profile within the interior ($r<R$) to a boundary layer solution near the edge ($r>R$) which joins smoothly to the temperature $T_\ice$ of the dense fuel.   
 \label{fig:Notation}}
 \end{figure}

Our analysis builds upon the well-known isobaric model~\cite{Atzeni_1984,Betti_2001,Springer_2019} which assumes the evolution of the hot spot radius $R(t)$ is slow in comparison to the sound crossing time.  This assumption is generally well satisfied in 1D simulations up through ignition, and implies that the hot spot temperature profile $T(r,t)$ is constrained by transport on fast time scales.   We also assume that the ion and electron temperatures are nearly in equilibrium, which is a good approximation prior to ignition.  The hot spot radius $R(t)$ is commonly defined to encompass most of the fusion reactivity.  Typically, this is specified by the location where the fusion reactivity per unit volume $Y_n(r)$ is some fraction of the central peak ($X \equiv Y_n(R)/Y_n(0)$).  For isobaric hot spots, the reactivity scales as $Y_n \propto \rho^2 T^n \propto T^{n-2}$, where $ n \approx 3.5$ at $T \sim 5$~keV.  Thus, defining $R(t)$ by a reactivity threshold is closely related to a threshold on the edge temperature $T_m(t) \equiv T(R(t),t)$ of the hot spot relative to the central temperature $T_o(t)$.   To aid the reader, a summary of our notation is shown in Fig.~\ref{fig:Notation}

\subsection{Hot Spot Power Balance}

To derive the hot spot power balance, we start with the total energy conservation equation
\begin{equation}
\frac{\partial}{\partial t} \left( \rho \varepsilon +\frac{\rho U^2}{2} \right) + \nabla \cdot \left[ {\bf U} \left( \rho \varepsilon  + \frac{\rho U^2}{2} \right)  + p{\bf U} \right] = S_{net} \;,
\label{eq:EnergyConservation}
\end{equation}
where $\rho$ is the mass density, $\varepsilon$ is the specific internal energy, $p$ is the pressure, ${\bf U}$ is the flow velocity and $S_{net} \equiv S_{fus} - S_{rad} - S_e$ is the net power per volume due to fusion heating ($S_{fus} \equiv S_\alpha + S_n$), radiation and thermal transport.  For subsonic regimes, the kinetic energy density is small in comparison to the internal energy ($U^2 \ll 2\varepsilon$) and some researchers \cite{Betti_2001,Zhou_2008} formally order in terms of $\delta^2 \equiv U_o^2/\varepsilon \propto (U_o/C_s)^2 \ll 1$, where $C_s$ is the sound speed and $U_o$ is a characteristic fluid velocity. Non-uniformities in the hot spot pressure enter as $\delta^2$ corrections to the isobaric approximation, together with a range of other difficulties.   For a hot spot approaching stagnation ($T\sim 5$~keV, $C_s \sim 600$~km/s, $U_o <200$~km/s) the isobaric approximation is quite good ($\delta^2 <0.1$).  Assuming spherical symmetry and retaining only the zeroth-order leading terms in the expansion, we can integrate Eq.~(\ref{eq:EnergyConservation}) over the time-evolving hot spot volume to obtain
\begin{equation}
\frac{d E}{dt} + 4 \pi R^2 \left[ p \frac{dR}{dt}  - \rho v_a \left( \varepsilon +\frac{p}{\rho} \right) \right]_{r=R(t)} = q_{net} \;,
\label{eq:EnergyConservation2}
\end{equation}
where $E \equiv \int \rho \varepsilon \, dV$ is hot spot energy, $v_a \equiv (dR/dt - U_r)$ is the ablation velocity, $U_r$ is the radial fluid velocity at the hot spot radius $R(t)$, and $q_{net} \equiv \int S_{net} \,dV$.  The terms in the bracket are evaluated at the edge of the hot spot, with the first corresponding to the standard compressional work.  The second term in the brackets corresponds to the enthalpy ($h \equiv \varepsilon + p/\rho)$ flux into the hot spot due to mass ablation.   Some researchers \cite{Betti_2001,Zhou_2008} explicitly assumed $v_a \ll dR/dt$ and neglected the enthalpy flux, while other researchers have assumed a cancellation of this term with a portion of the heat loss (see Sec.~\ref{sec:massAblation}).  While this may be justified in the limit $T_\ice \rightarrow 0$,  it is not correct for finite $T_\ice$ where $v_a \sim dR/dt$ near stagnation (see Sec.~\ref{hotspotEvolution}).  Our theoretical analysis and comparisons with radiation-hydrodynamic simulations demonstrate that the enthalpy flux is comparable to the other terms in Eq.~(\ref{eq:EnergyConservation2}) and thus plays a key role in the ignition physics.   Furthermore, by evaluating each term in Eq.~(\ref{eq:EnergyConservation2}) from the simulations, we demonstrate that the overall energy balance is in good agreement with the assumed ordering ($\delta^2 \ll1$). 


While Eq.~(\ref{eq:EnergyConservation2}) is valid for arbitrary equation-of-state (EOS), to proceed further we consider a gamma-law EOS ($p = (\gamma-1) \rho \varepsilon$) where $\gamma \equiv C_P/C_V=5/3$ is the ratio of specific heats at constant pressure/volume.  Assuming the hot spot is isobaric ($p(r,t) =P(t))$, and using mass conservation ($dM/dt = 4 \pi R^2 \rho_m v_a$)  we can re-write Eq.~(\ref{eq:EnergyConservation2}) as 
\begin{equation}
\frac{d E}{dt} +  P \frac{d V}{dt} - h \frac{dM}{dt} = q_{net} \equiv q_{fus}- q_{rad} - q_e \;,
\label{eq:EnergyConservation3}
\end{equation}
where $E$, $M$, $V$ and $P$ are the hot spot energy, mass, volume and pressure, $q_{fus}$ is the fusion power ($\alpha + n$) deposited within the hot spot, $q_{rad}$ and $q_e$ are the radiative and conduction losses, $h \equiv C_P T_m$ is the enthalpy at the edge and $C_P$ is the specific heat at constant pressure.  The enthalpy term in Eq.~(\ref{eq:EnergyConservation3}) accounts for both the flux of internal energy due to the ablative inflow and the work performed on the hot spot by this inflow.   Since $h dM/dt$ is positive, if $q_{net}>0$ near stagnation the hot spot energy will increase rapidly.  However, this well-known \cite{Atzeni_2004} static ignition threshold ($q_{fus}>q_{rad} + q_e$) does not imply the temperature will increase, since $q_{net}$ is distributed over the increasing hot spot mass. 

To understand the influence of the enthalpy flux on the mass averaged hot spot temperature $\bar{T} \equiv E/(C_V M)$, one may re-write Eq.~(\ref{eq:EnergyConservation3}) as
\begin{equation}
C_V \frac{d {\bar T}}{dt}  + \frac{P}{M} \frac{dV}{dt} + C_V {\bar T} \left( 1-\gamma \frac{T_m}{{\bar T}} \right) \frac{1}{M}\frac{dM}{dt} = Q_{net}\;,
\label{eq:TemperatureEvolution1}
\end{equation}
where $Q_{net} \equiv Q_{fus} - Q_{rad} - Q_e$.  Here the radiative $Q_{rad}$ and conduction power $Q_e$ are also normalized to the hot spot mass, and $Q_{fus} \equiv f_{\alpha} Q_{\alpha}+f_{n} Q_{n}$ where $Q_{j}=q_{j}/M$ is the total power per mass from fusion products ($j = \alpha, n$) and $f_j$ is the fraction deposited within the hot spot ($ 0 \le f_j \le 1$).  Notice that the $dM/dt$ term in Eq.~(\ref{eq:TemperatureEvolution1}) includes a contribution arising from $dE/dt$ in Eq.~(\ref{eq:EnergyConservation3}) which competes with the enthalpy flux.  If $R(t)$ is chosen to encompass most of the fusion reactivity, then $T_{m} <{\bar T}/\gamma$ and mass ablation has a cooling influence (see Appendix~\ref{MassAppendix} for discussion of temperature profile).   

\subsection{Mass Ablation Rate}
\label{sec:massAblation}
The mass ablation into the hot spot is driven by the conduction loss and by $\alpha$-particles deposited in the dense fuel.
From our new analysis (see Appendix~\ref{MassAppendix}), the hot spot mass evolution is 
 \begin{equation}
C_V {\bar T} \frac{1}{M} \frac{dM}{dt} \approx {\mathcal A} \left[ Q_\alpha (1-f_\alpha) + Q_e \right] \;,
\label{eq:MassAblation}
\end{equation}
where ${\mathcal A} = 1 +  (\rho_m/\bar{\rho})/\ln (T_m/T_\ice)$ is a dimensionless factor arising from a boundary layer analysis,  $\rho_m$ is the edge density, ${\bar \rho} \equiv M/V$ and $T_\ice$ is the temperature of the dense fuel outside the hot spot. 

 In the limit $T_\ice \rightarrow 0$, then ${\mathcal A} = 1$ and Eq.~(\ref{eq:MassAblation}) is identical to previous work \cite{Atzeni_1984,Hurricane_2019,Hurricane_2019B,Hurricane_2021,MacLaren_2021}.   Furthermore,  by grouping the work done by the ablative inflow with $P dV/dt$,  one may use Eq.~(\ref{eq:MassAblation}) to eliminate the remaining portion of $h dM/dt$ in Eq.~(\ref{eq:EnergyConservation3}).  The resulting power balance is identical to previous work \cite{ Atzeni_2004, Betti_2001} but still includes a $dM/dt$ contribution within the $dE/dt$ term, which is typically neglected in the ignition threshold \cite{Hurricane_2019,Hurricane_2021}.   Since our analysis demonstrates that ${\mathcal A} > 1$ for $T_\ice > 0$, this cancellation does not occur as previously assumed. The validity of Eq.~({\ref{eq:MassAblation}) is independently confirmed by numerically solving the underlying nonlinear diffusion equation from which ${\mathcal A}$ is derived (see  Appendix~\ref{MassAppendix}) and also by comparing with radiation-hydrodynamic simulations discussed below. For capsules approaching ignition, the enhancement factor is in the range $ 2 \lesssim {\cal A} \lesssim 4$.
 
 \subsection{Ignition Threshold} 
The compressional heating of the hot spot is negligible for a brief time interval near stagnation.  In addition, the neutron heating within the hot spot is also small near stagnation due to the modest areal density of the hot spot ($\rho R \sim 0.3\; {\rm g/cm^2}$) at this time.  However, since the total areal density including the dense fuel is much larger ($\rho R \sim 0.85 \;{\rm g/cm^2}$), the neutron heating within the dense fuel is significant (see Sec.~\ref{ignitionThreshold}).  This heating will increase the temperature of dense fuel and the mass ablation rate into the hot spot, but does appear directly in the hot spot power balance.  To derive an ignition threshold valid near stagnation, we use Eq.~(\ref{eq:MassAblation}) to eliminate $dM/dt$ in Eq.~(\ref{eq:TemperatureEvolution1}) and neglect both compressional and neutron heating within the hot spot to obtain
\begin{equation}
C_V \frac{d {\bar T}}{dt} \approx Q_\alpha  \left[f_\alpha -D \left(1-f_\alpha \right) \right] - Q_{rad} - Q_e (1+D) \;,
\label{eq:StagnationTemperature}
\end{equation}
where $D \equiv {\mathcal A} (1-\gamma T_m/{\bar T})$ is a dimensionless parameter that depends on the thermodynamic state of the dense fuel and controls the net influence of mass ablation.  Recall that $T_m/{\bar T}$ is determined by the reactivity threshold $X$ defining the hot spot radius (see Appendix~\ref{MassAppendix}).  Assuming $X \approx 0.1$ and $\gamma=5/3$, one may roughly estimate  $T_m/\bar{T} \approx 0.45$ and $ D \approx 0.25 {\mathcal A}$ for a hot spot approaching ignition temperature (${\bar T} \approx 5$~keV).   Well before stagnation, the dense fuel is cold resulting in ${\mathcal A}\approx 1$ and $D \sim 0.25$.  Approaching stagnation, heating within the dense fuel increases $T_\ice$ resulting in ${\mathcal A} \sim 2 $ ($D \sim 0.5$), while shortly after stagnation this increases further to ${\mathcal A} \sim 4 $ ($D \sim 1$). 

The ignition threshold is estimated by requiring $d{\bar T}/dt>0$ in Eq.~(\ref{eq:StagnationTemperature}) and introducing approximate analytic scalings (see Appendix~\ref{IgnitionAppendix}) following previous work \cite{Atzeni_1984}.  The new threshold is a function of ${\bar T}$, $\rho R$ and $D$ which depends primarily on $T_\ice$ of the dense fuel.   In the limit $D=0$, Eq.~(\ref{eq:TemperatureEvolution1}) reduces to previous work \cite{Atzeni_1984} which simply corresponds to $q_{fus}>q_{rad} + q_e$.   As shown in Eq.~(\ref{eq:EnergyConservation3}), this criterion still has important physical significance, since it determines the stagnation conditions where fusion heating will amplify the hot spot energy.  Ignition corresponds to a rapid increase in the temperature $d{\bar T}/dt>0$ which as shown in Eq.~(\ref{eq:StagnationTemperature}) is more stringent due to the mass ablation $dM/dt$ terms.  

\section{Radiation-Hydrodynamic Simulations}
\label{radhydro}

\begin{figure}[t]
\includegraphics[width=\linewidth]{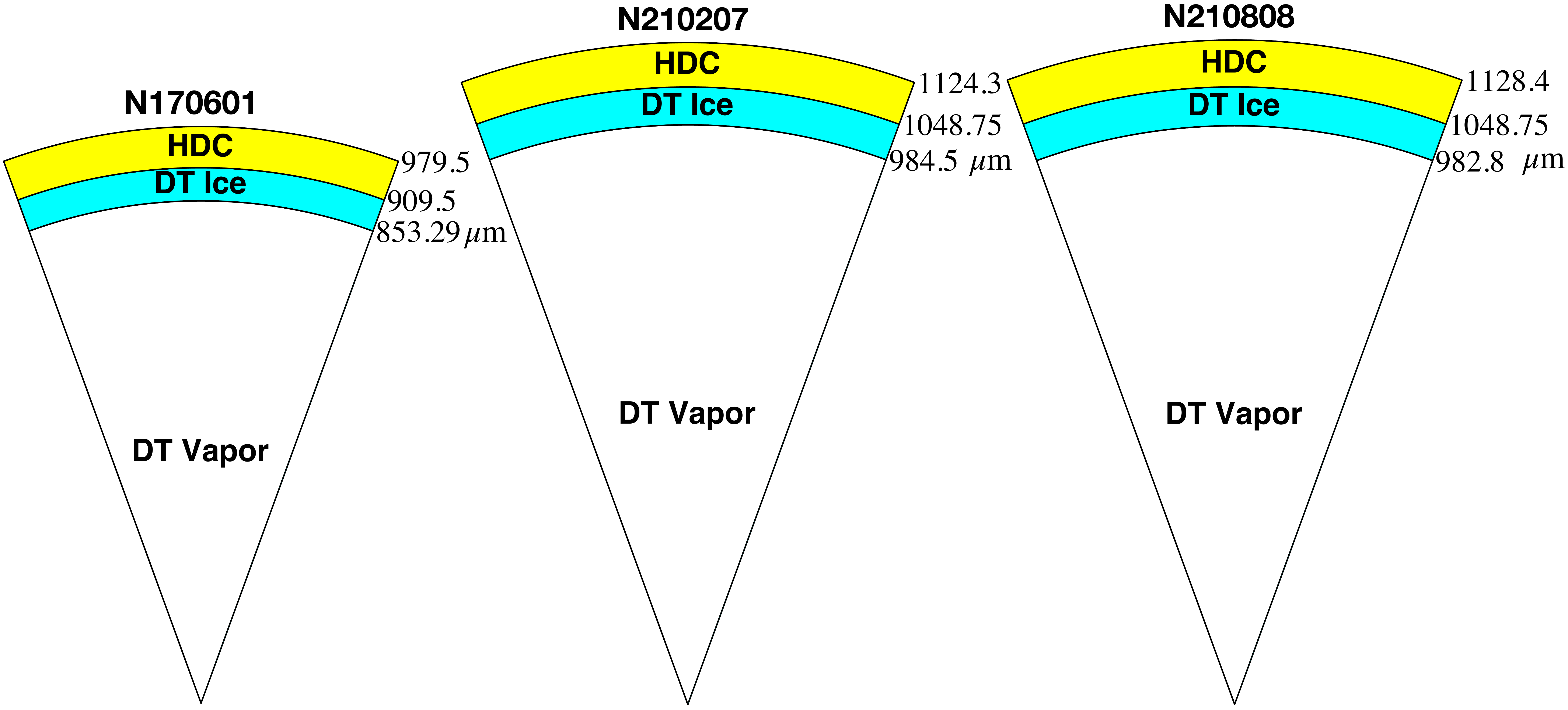}
\caption{Capsule geometries for three recent NIF experiments. \label{fig:Pie}}
\end{figure}

We consider two nearly identical capsules shown in Fig.~\ref{fig:Pie}, N210207 which entered the burning plasma regime (0.17 MJ) but did not ignite \cite{Zylstra_2022} and N210808 which is the first capsule to successfully ignite (1.39 MJ) on the NIF \cite{Hurricane_2022}.  Of relevance to 1D implosion physics, N210808 employed smaller laser entrance holes to increase the radiation drive and a slightly longer pulse to reduce the coast time \cite{Hurricane_2022C} between the end of the pulse and stagnation.  While the excellent quality of the N210808 capsule also likely contributed, 1D simulations show a dramatic increase in the yield ($\sim13\times$) in comparison to N210207.  To further illustrate the ignition physics, we also consider a somewhat smaller HDC capsule N170601 \cite{Pape_2018} which demonstrated good performance (48 kJ).   Simulations were performed for each of these capsules using the xRAGE radiation-hydrodynamics code \cite{Gittings_2008,Haines_2017} for modeling indirectly driven implosions as outlined in Ref.~\cite{Haines_2017}.  The simulations employ adaptive mesh refinement (AMR) with $1/4\,\mu {\rm m}$ resolution in the ablator and $(1/32) \,\mu {\rm m}$ in the ice layer to more accurately track the mass ablation.  Simulations follow the post-shot modeling strategy described in Ref.~\cite{Clark_2013} and are driven using a frequency-dependent x-ray boundary source (FDS) derived from hohlraum simulations performed in the HYDRA radiation-hydrodynamics code~\cite{Marinak_2001}.  These drives are tuned to match experimentally measured shell trajectories, shock timing, and bang time as measured by convergent ablator, keyhole, and layered DT experiments, respectively.
 The simulations employed multi-group diffusion for the radiation transport (60 energy groups), electron and ion thermal conduction, and three-temperature ($T_i$, $T_e$, $T_r$) burn physics~\cite{Haines_2022} with charged-particle transport for the $\alpha$-particle deposition.  Neutron heating was treated within a single-scatter approximation \cite {Lindl_2004}.  Key experimental and simulated quantities are summarized in Table~\ref{Table:KeyStuff}. 

\begin{table}[t]
\caption{Summary of experimental \cite{Pape_2018,Zylstra_2022,Hurricane_2022} and simulated quantities for the capsules shown in Fig.~\ref{fig:Pie}.  The experimentally measured temperature $\langle T_{i} \rangle$ corresponds to the burn weighted ion temperature obtained from DT neutrons.\label{Table:KeyStuff}}
\begin{center}
\begin{tabular}{cc|c|c|c|}
    Quantity &&  N170601 & N210207 & N210808  \\ 
\hline
Bang time        & (exp)   &  $8.16 \pm 0.03$ &  $9.11 \pm 0.02$  &  $9.28\pm0.07$    \\
   (ns)      & (sim)  & 8.17 &  9.14  &  9.16   \\
\hline                         
Burn width & (exp)   &  $165\pm30$ &  $120\pm20$  &  $89\pm15$    \\
         (ps)     & (sim)  & 108 &  112  &  43    \\
\hline                      
$\langle T_{i} \rangle$       & (exp)   &  $4.5 \pm 0.12$ &  $5.6 \pm 0.1$  &  $10.9\pm0.4$    \\
   (keV)      & (sim)  & 6.2 &  6.4  &  13.7   \\
 \hline 
DT  neutrons &  (exp)   &  $1.7 \pm 0.02$ &  $6.1 \pm 0.2$  &  $48\pm 1$    \\
      ($10^{16}$)    &  (sim)    &  8.9 &  17  &  221    \\
 \hline 
 $R(t_{stag})\;\;(\mu {\rm m})$ &  (sim)    & 36 & 44.6  &  43.7    \\
 $M(t_{stag})\;\;(\mu {\rm g})$ &  (sim)     & 15 & 24  &  30    \\
  $E(t_{stag})\;\;({\rm kJ})$ &  (sim)     & 9.4 & 13.8  &  18.7    \\
  $\bar{T}(t_{stag})\;\;({\rm keV})$ &  (sim)     & 4.9 & 4.7  &  5.3    \\
  $v_a(t_{stag})\;\;(\mu {\rm m/ns})$ &  (sim)    & 79 & 88  &  100    \\
\hline                      
 
 \end{tabular}
\end{center}
\label{default}
\end{table}%

\subsection{Hot Spot Evolution}
\label{hotspotEvolution}
To illustrate the hot spot evolution, we choose the reactivity threshold $X = 0.1$ to define the hot spot radius $R(t)$, which is consistent with previous work.    Note that Eqns.~(\ref{eq:EnergyConservation3})-(\ref{eq:TemperatureEvolution1}) are valid for any choice of $R(t)$, as confirmed by comparison with simulations.  The boundary layer analysis for the mass ablation in Eq.~(\ref{eq:MassAblation}) requires $T_o \gg T_m \gg T_\ice$, and thus there is freedom to choose $X$ over some range.  While all hot spot quantities ($\rho R$, $\bar{T}$, $D$) will vary with the threshold defining $R(t)$, the ignition threshold determined by Eq.~(\ref{eq:StagnationTemperature}) is valid for any reasonable choice, and our conclusions regarding ignition are not sensitive to this choice.

\begin{figure*}
 \includegraphics[width=1.01\linewidth]{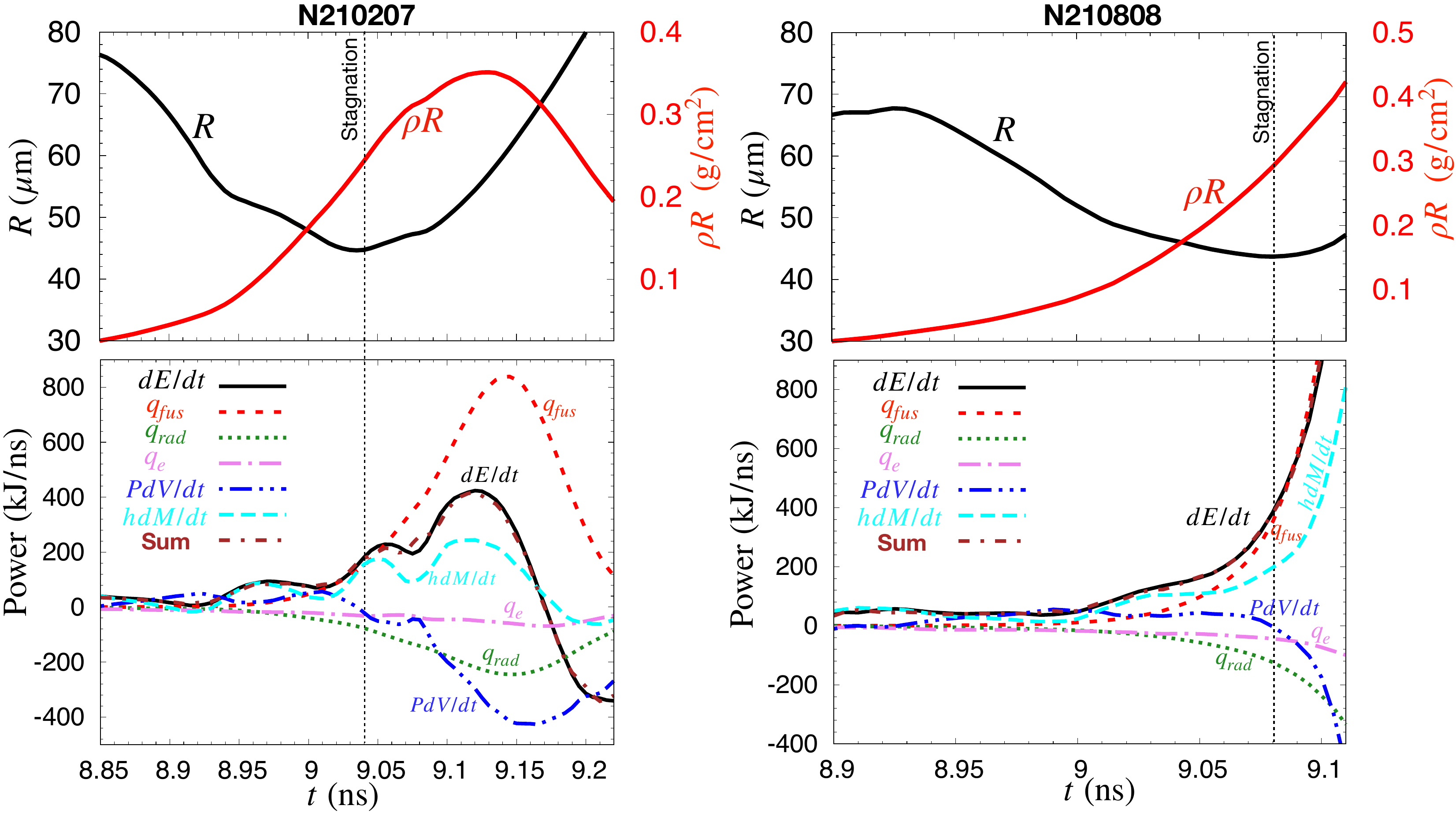}
 \caption{Hot spot radius $R(t)$ and areal density $\rho R \equiv \int_0^R \rho dr$ (top) and power balance (bottom) for the largest two capsules in Fig.~\ref{fig:Pie}. \label{fig:RadiusEnergy}}
 \end{figure*}  
 
The resulting evolution for the hot spot radius $R(t)$ and areal density $\rho R \equiv \int_0^R \rho dr$ are illustrated in the top panels of Fig.~\ref{fig:RadiusEnergy} for the larger two capsules.  To evaluate the terms in Eq.~(\ref{eq:EnergyConservation3}), the ice layer was divided into 25 subregions and the full power balance was saved for each of these regions.  Each term in Eq.~(\ref{eq:EnergyConservation3}) is computed independently by summing over the regions with $r \le R(t)$, as illustrated in the bottom panels of Fig.~\ref{fig:RadiusEnergy}.    As a consistency check, the summation over all the heating terms is compared with actual $dE/dt$ evolution. The overall balance is well-satisfied (deviations $<5\%)$ which confirms the approximations that led to Eq.~(\ref{eq:EnergyConservation3}) and the correctness of our methodology for evaluating each term in the simulations.  

Notice that the enthalpy flux $h dM/dt$ in Fig.~\ref{fig:RadiusEnergy}  is a significant contribution to $dE/dt$ for both capsules, both during the early formation of the hot spot and approaching stagnation.   To understand the origin of this term, the evolution of the ablative inflow velocity $v_a \equiv (dR/dt - U_r)$ is shown in Fig.~\ref{fig:dRdt} for the N210808 capsule.  Approaching stagnation, the inflow into the hot spot is $v_a \sim 100 \mu {\rm m/ns}$.   The inflow velocities for the other capsules are also of similar magnitude near stagnation (Table~\ref{Table:KeyStuff}).  While previous authors \cite{Betti_2001,Zhou_2008} have assumed $v_a \ll dR/dt$, this is clearly not the case in Fig.~\ref{fig:dRdt}. Shortly after stagnation ($U_r \approx 0$) as the hot spot approaches ignition, the inflow $v_a$ is driven entirely by the outward expansion of the heat front $dR/dt$ which depends upon the temperature $T_\ice$ of the dense fuel (see Appendix~\ref{MassAppendix}).   As the hot spot ignites, the ablative inflow increases ($v_a > 200 \mu {\rm m/ns}$) due to the increasing heat loss into the dense fuel (see Eq.~(\ref{eq:MassAblation})).   Clearly, the expansion of the heat front $dR/dt$ is directly responsible for the enthalpy flux into the hot spot.   As demonstrated in Appendix~\ref{MassAppendix}, this physics is only described properly by considering finite $T_\ice$ of the dense fuel.   While the velocities shown in Fig.~\ref{fig:dRdt} correspond to a reactivity threshold $X=0.1$ to define $R(t)$,  the results are not sensitive to this choice.  In particular, within the threshold range $X=0.05 - 0.15$, the inferred velocities typically differ by only a few percent, with maximum differences of $ \pm 5 \%$  as the ablation rate increases after stagnation.   

\begin{figure}
 \includegraphics[width=\linewidth]{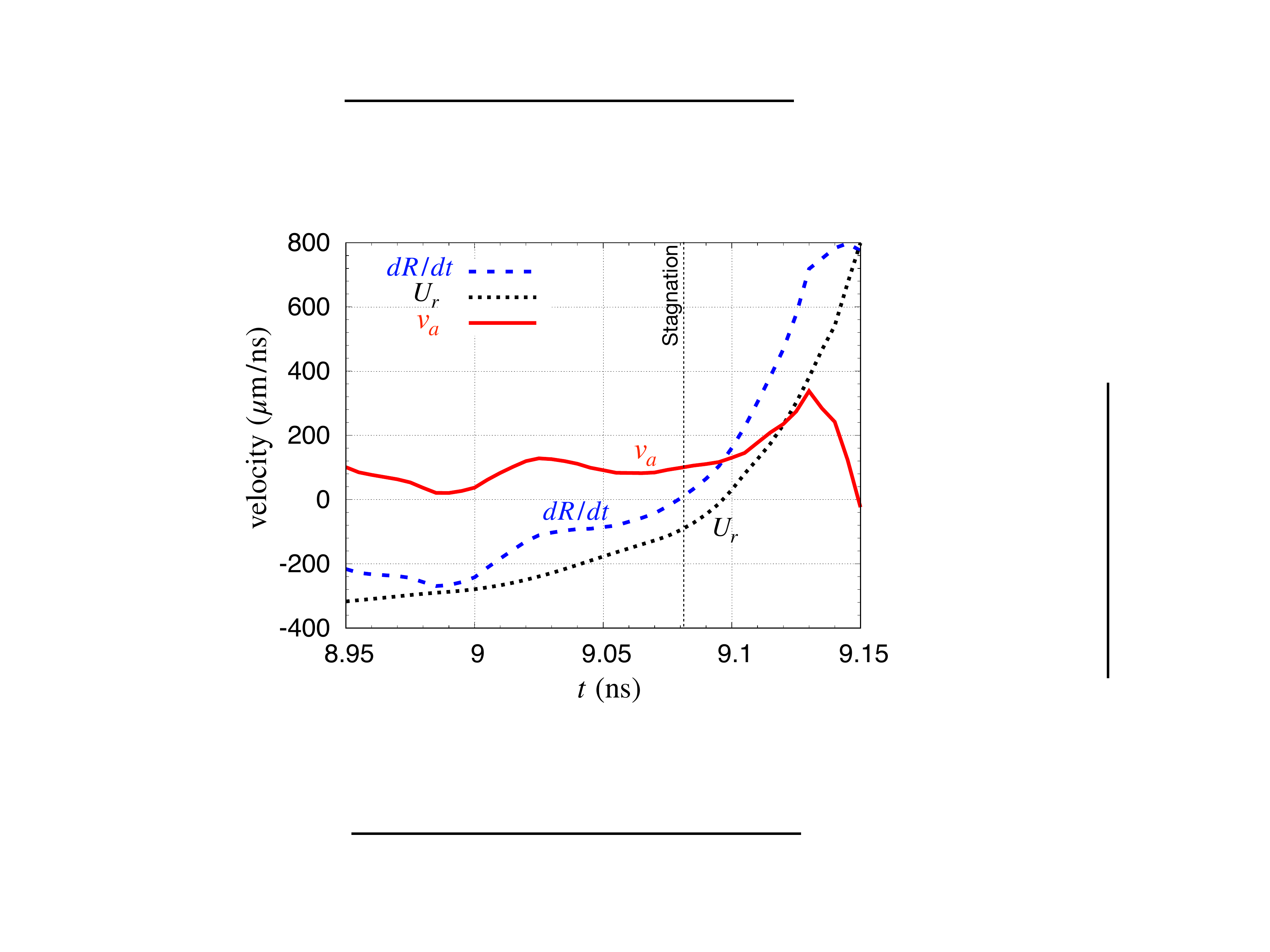}
 \caption{Evolution of the fluid velocity $U_r$ at the edge of the hot spot ($r=R(t)$) and the heat front velocity $dR/dt$ for N210808.  The mass ablation velocity $v_a \equiv dR/dt - U_r$ varies weakly near stagnation ($\sim 100 \mu {\rm m/ns}$), but then increases rapidly as the hot spot ignites. \label{fig:dRdt}}
 \end{figure}

Shortly before stagnation, N210207 satisfies $q_{fus} > q_{rad} + q_e$, leading to a significant increase in $E$ as expected from Eq.~(\ref{eq:EnergyConservation3}).  Since the net heating is distributed over an increasing mass, it becomes difficult to increase the temperature and the capsule fails to ignite.  As hot spots enter the regime of strong $\alpha$-heating, it is easier to increase energy $E$ ($M$, $\rho R$) than the temperature ${\bar T}$.   This is also apparent in the bottom right panel of Fig.~\ref{fig:RadiusEnergy} for N210808, where the thermonuclear runaway begins in $E$ well before stagnation.  This leads to a rapid increase in mass ablation driven by $\alpha$-particles deposited in the dense fuel.  While this increases the $\rho R$ of the hot spot, these inflows inhibit the temperature increases for another 60 ps, until the entire dense fuel is heated.    As shown in Fig.~\ref{fig:Tprofile}, the temperature profile expected for isobaric hot spots (see Appendix~\ref{MassAppendix}) remains accurate, with deviations appearing within $\sim 25$~ps of peak burn as the temperature grows explosively.  

\begin{figure}
 \includegraphics[width=\linewidth]{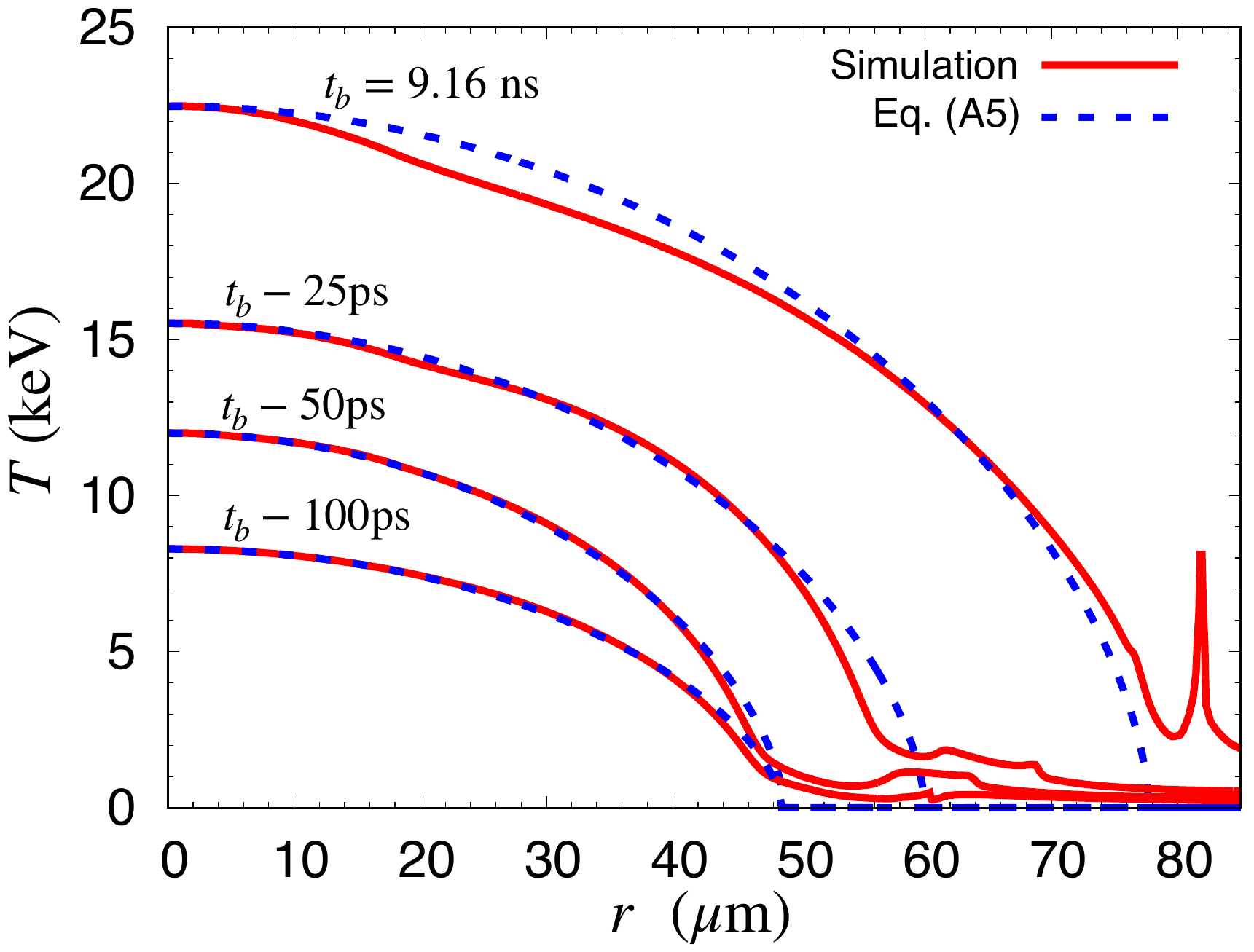}
 \caption{Comparison of the ion temperature profile for N210808 obtained from simulation versus the analytic form in Eq.~(\ref{eq:Tprofile}) using $\beta=0.67$.   In the simulation, stagnation occurs at $\sim 9.08$ ns and peak burn occurs at $t_b = 9.16$ ns. \label{fig:Tprofile}}
 \end{figure}
  \begin{figure*}[t]
 \includegraphics[width=\linewidth]{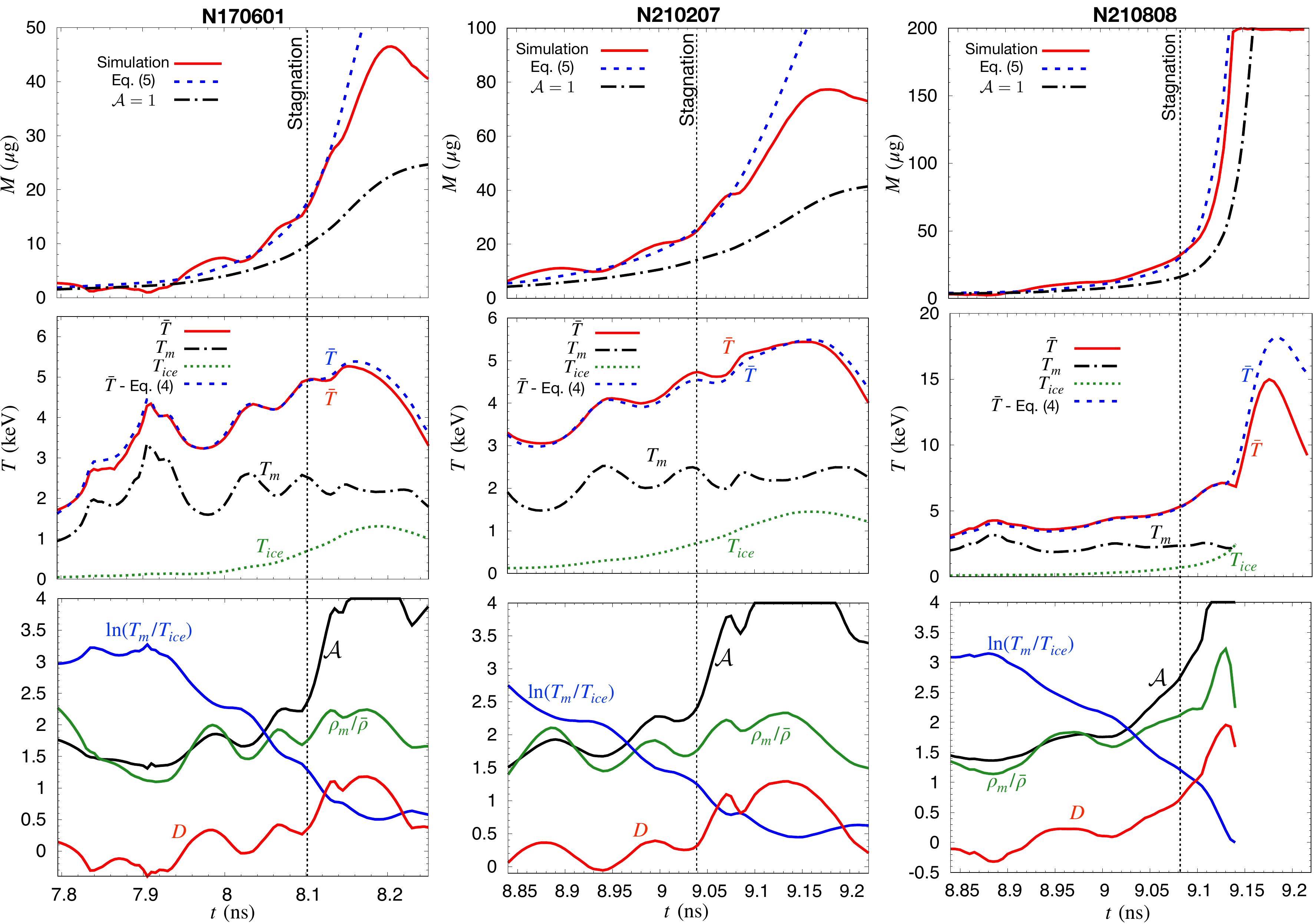}
 \caption{Evolution of hot spot mass (top), temperature (middle) and key dimensionless parameters (bottom) for the three capsules shown in Fig.~\ref{fig:Pie}. \label{fig:MassTemp}}
 \end{figure*}

 While in Fig.~(\ref{fig:RadiusEnergy}) the $hdM/dt$ term was evaluated directly from the simulations,  one can check the new $dM/dt$ prediction by integrating Eq.~(\ref{eq:MassAblation}) using $Q_e$, $Q_\alpha$ and $f_\alpha$ from the simulation.  The ice temperature $T_\ice$ is estimated by the mass-averaged temperature of the fuel outside the hot spot ($r>R(t)$).  As shown in the top panels of Fig.~\ref{fig:MassTemp},  the new model for the hot spot mass (blue dashed)  is in reasonable agreement with the actual hot spot mass in the simulation (red), thus confirming the factor ${\mathcal A}$ arising from  the boundary layer analysis.   Deviations are largest at late time, where the increasing $T_\ice$  (middle panel) leads to the condition $T_m/T_\ice <2$  where the theory breaks down.  The previous expression for mass ablation is slower by the factor ${\mathcal A}$, which is shown in the bottom panel along with other dimensionless parameters.  Since both $\rho_m/\bar{\rho}$ and $T_m$ are tied to the reactivity threshold, they do not vary widely and the increase shown in ${\mathcal A}$ is primarily due to increasing $T_\ice$. 
 
As another consistency check, Eq.~(\ref{eq:TemperatureEvolution1}) for the hot spot temperature was integrated in time, using the hot spot diagnostic to evaluate each term from the simulation.  As shown in the middle panel of Fig.~\ref{fig:MassTemp}, the model prediction (blue dashed) is in good agreement with the actual temperature (red) for each simulation.  This confirms the validity of the approximations (isobaric, gamma law, etc) leading to Eq.~(\ref{eq:TemperatureEvolution1}).   Notice that  for N210207, the hot spot temperature only increases by $0.7$~keV from stagnation to peak burn, while $E$ increases by $\sim 3\times$, due to the enthalpy flux (see Fig 2).  This is also evident in N210808, where the  runaway in $E$ begins near $t=9.06$~ns with temperature $T\approx 5.5$~keV, but over the next 70~ps the temperature only increases to 7~keV, as the hot spot consumes the entire ice layer $M \approx 200\;{\mu \rm g}$ leading to a rapid increase in $\rho R$.   At this point, the temperature increases explosively $T\approx 7 \rightarrow 15$~keV, producing most of the yield. 

\subsection{Ignition Threshold}
\label{ignitionThreshold}

The new ignition threshold in Eq.~(\ref{eq:StagnationTemperature}) corresponds to a surface in the parameter space ($\rho R$, ${\bar T}$, $D$) as shown in the top panel of Fig.~\ref{fig:Ignition}  (Multimedia view).  For each capsule, the hot spot evolution corresponds to a trajectory within this space and ignition occurs when a trajectory crosses this surface.   One may also visualize the ignition space using the 2D projection ($\rho R$, ${\bar T}$) shown in the bottom panel, where the contours of $D$ are computed from Eq.~(\ref{eq:StagnationTemperature}).   While this illustrates the hot spot evolution in the standard ($\rho R$, ${\bar T}$) space, it is important to remember that $D$ is also evolving with $T_\ice$ of the dense fuel.  
\begin{figure}
\includegraphics[width=\linewidth]{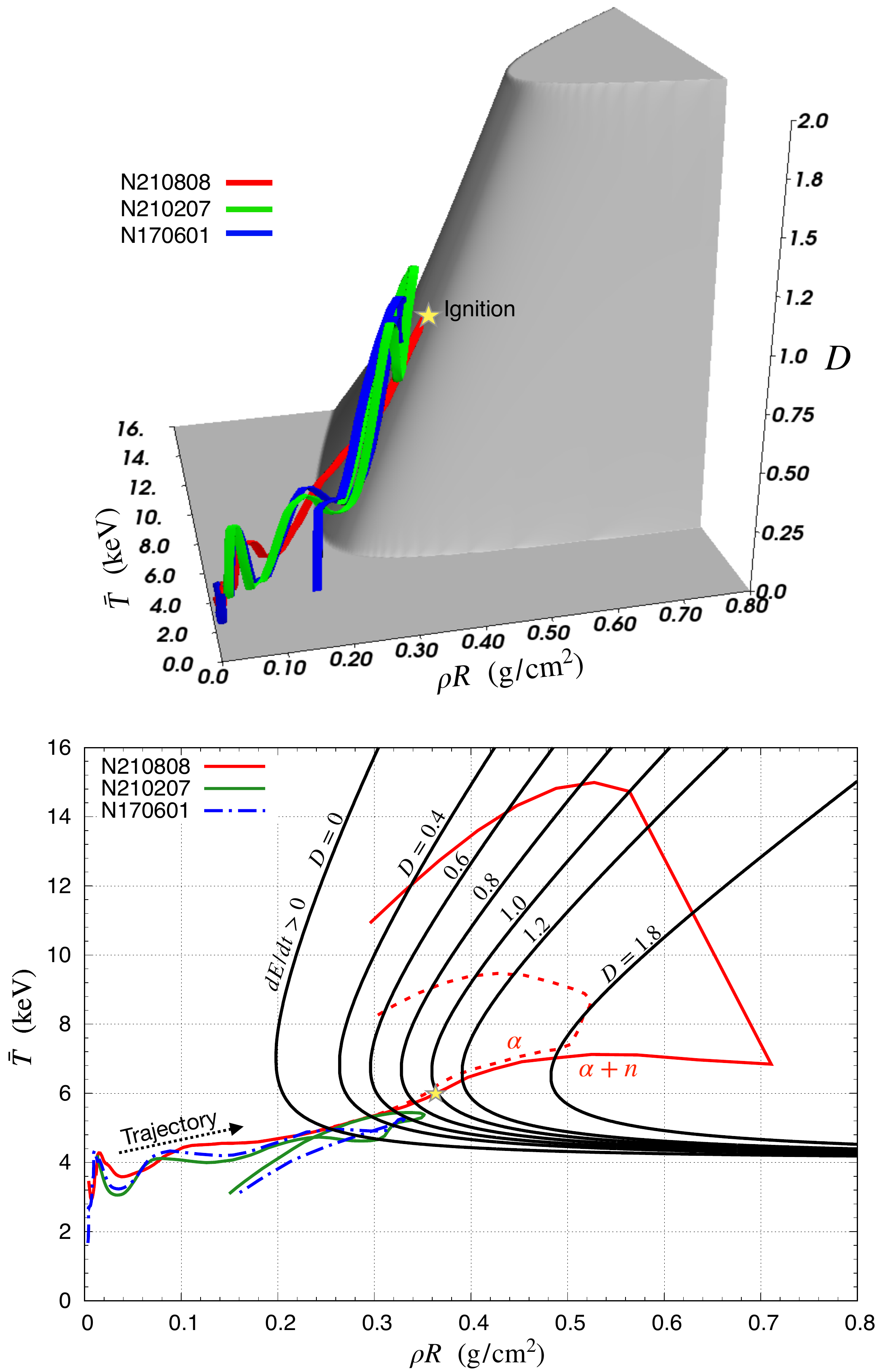}
\caption{The ignition threshold defined by Eq.~(\ref{eq:StagnationTemperature}) corresponds to a surface in the 3D parameter space ($\rho R$, ${\bar T}$, $D$) shown in the top panel.  The evolution of the hot spot trajectories are shown for the capsules in Fig.~\ref{fig:Pie}.  The trajectory for N210808 crosses the ignition surface (yellow star), while the other capsules cross into $dE/dt>0$ region, but fail to reach the ignition surface.  Bottom panel shows the 2D projection with contours of $D$ indicated.   For N210808,  one simulation included both $\alpha$ and neutron heating (solid red) while the other included only $\alpha$-heating (dashed red).   See the online supplement for a 3D animation of the ignition surface and hot spot trajectories. (Multimedia view)  \label{fig:Ignition}}   
\end{figure}

 Shortly before stagnation, the  $N210207$ trajectory in Fig.~\ref{fig:Ignition} crosses the $D=0$ boundary corresponding to $dE/dt >0$ in Eq.~(\ref{eq:EnergyConservation3}).  The net heating $q_{fus} > q_{rad}+q_e$ drives additional ablation into the hot spot increasing the areal density $\rho R = 0.25 \rightarrow 0.34 \; {\rm g/cm^2}$ near peak burn, while also increasing $D = 0.4 \rightarrow 1.2$  over this interval.   As a result, the hot spot trajectory ($\rho R$, ${\bar T}$, $D$) never crosses the ignition surface shown in the upper panel.   The trajectory for N210808 crosses the $D=0$ curve shortly before stagnation, leading to a rapid enhancement in $E$, $M$, $\rho R$ and $D$.   This allows the trajectory to cross the ignition surface near  ($\rho R \approx 0.37 \;{\rm g/cm^2}$, ${\bar T} \approx 6.1$~keV, $D\approx 1$) as shown in Fig.~\ref{fig:Ignition} by the yellow star.   
   
 To illustrate the influence of neutron heating, two trajectories are shown for the $N210808$ capsule in the bottom panel of Fig.~\ref{fig:Ignition}.   In the simulation with both $\alpha$ and neutron heating (solid red) the hot spot areal density is driven to $\rho R \approx 0.71 \;{\rm g/cm^2}$, while in the simulation with only  $\alpha$-heating (dashed red) the maximum areal density $\rho R \approx 0.52 \;{\rm g/cm^2}$ is significantly lower.   Since neutron heating within the hot spot is negligible for $\rho R <0.35 \;{\rm g/cm^2}$ it is typically neglected in ignition theory.  However, the larger areal density within the dense fuel ($\rho R \sim 0.8 \;{\rm g/cm^2}$) results in significant neutron heating, which increases $T_\ice$ and the mass ablation into the hot spot. Paradoxically, the additional heating in the dense fuel results in slightly lower ${\bar T}$ in the hot spot as the heat front expands.  By driving the hot spot to larger $\rho R$, the temperature increase at late time is much faster and involves the entire fuel mass.  

While the influence of neutron heating on burn propagation has been considered in previous simulation studies \cite{Gauthier_2004}, the impact on the ignition threshold has not been appreciated. To illustrate the relative importance of various heating terms within the dense fuel, the power balance is shown in Fig.~\ref{fig:IcePower} for N210808.  At early time, the compressional work and thermal conduction are the dominant heating terms.  As expected, the conduction loss from the hot spot (see Fig.~\ref{fig:RadiusEnergy}) is entirely absorbed in the dense fuel.  At $\sim 40$~ps prior to stagnation, radiative heating (inverse bremsstrahlung absorption) overtakes thermal conduction and later becomes the dominant heating term near stagnation.   Surprisingly, the radiative heating power is up to $\sim 40$\% of the emitted radiation from the hot spot shown in Fig.~\ref{fig:RadiusEnergy}.  However, while conduction is deposited locally at the inner edge of dense fuel, the radiative heating is volumetric in nature.  Near stagnation time, the neutron heating is only slightly smaller than radiative heating.  As the capsule ignites, neutron heating becomes dominant and the entire dense fuel is rapidly absorbed into the hot spot.   
 \begin{figure}
 \includegraphics[width=\linewidth]{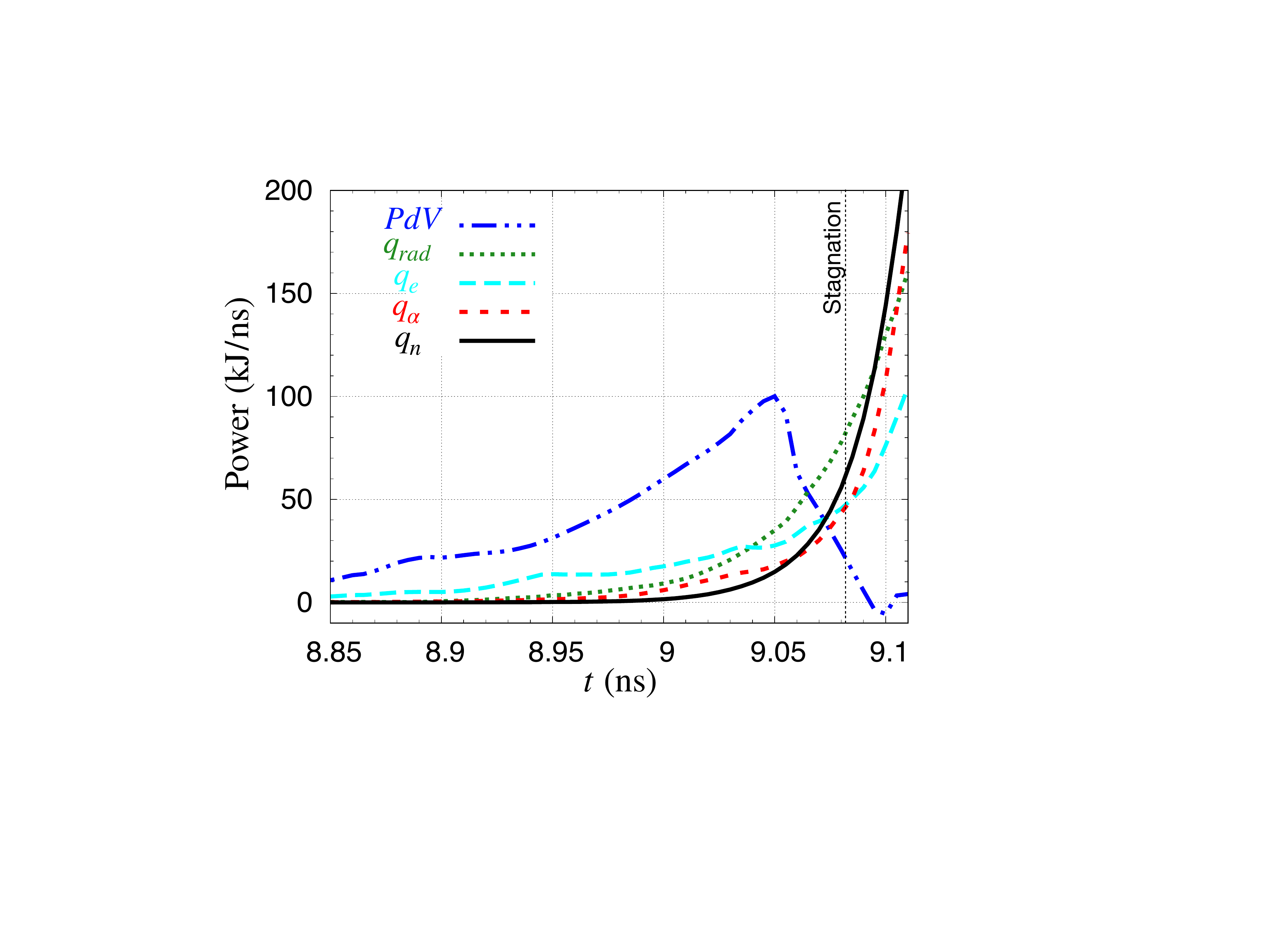}
 \caption{Heating terms for the dense fuel region for N210808.  \label{fig:IcePower}}
 \end{figure}

\section{Summary} 
\label{summary}
This work builds upon previous isobaric models for ignition \cite{Atzeni_1984,Betti_2001,Springer_2019}, with several important differences.  First, the previous expression for mass ablation into the hot spot is only valid for $T_\ice \rightarrow 0$, where the isobaric model features a singularity ($\rho \rightarrow \infty$) at the edge.  Relaxing this condition allows the heat front to propagate and the resulting mass ablation is up to  ${\mathcal A} \sim 4\times$ faster, depending on  $T_\ice$.   Second, our theory retains all $dM/dt$ terms in the power balance, including the enthalpy flux into the hot spot. This term acts to increase the hot spot temperature, while the $dM/dt$ term arising from $dE/dt$ corresponds to a redistribution of the net heating over the increasing hot spot mass.  The net influence on the temperature evolution is determined by the competition between these terms and is regulated by the dimensionless parameter $D \equiv {\mathcal A} (1-\gamma T_m/{\bar T})$, which increases strongly with $T_\ice$ of the dense fuel.  

As the hot spot approaches stagnation conditions, there are two important thresholds to consider.  First, if the net heating is positive $q_{fus} > q_{rad}+q_e$, the hot spot energy $E$ will grow rapidly as shown by Eq.~(\ref{eq:EnergyConservation3}) and observed in the simulations. 
This corresponds to the previous isobaric ignition threshold \cite{Atzeni_2004} (see $D=0$ curve in Fig.~\ref{fig:Ignition}).   While $\alpha$-heating in this regime will increase $M$ and $\rho R$, the hot spot temperature may be flat or even decreasing due to the increasing heat capacity of the hot spot driven by $dM/dt$.  Since this physics is included in Eq.~(\ref{eq:StagnationTemperature}), the new threshold ($d {\bar T}/dt>0$) implies a rapid increase in the fusion reactivity, which is the true hallmark of ignition.  
 
This analysis represents a fundamental shift in the understanding of layered implosions, since the ignition threshold is a surface in ($\rho R$, ${\bar T}$, $D$) parameter space.  Hot spots with similar ($\rho R$, ${\bar T}$) may perform very differently, depending upon $T_\ice$ of the dense fuel. As demonstrated in Fig.~\ref{fig:Ignition}, this implies that neutron heating within the dense fuel plays an important role in the ignition and burn propagation.  Other heating mechanisms, such as absorption of radiation emitted from the hot spot, will likely play a similar role.  The fraction of radiation deposited within the dense fuel may depend on high-Z impurities within the ablator, which are commonly included and which may mix into the ice during the implosion.  Other types of designs feature high-Z layers \cite{MacLaren_2021} at the inner ablator surface (near ice), which is expected to trap radiation and increase the temperature of the dense fuel.  Depending upon $T_\ice$, there is a significant portion of the parameter space where ignition is inaccessible within our new analysis. For example, even with $\rho R \approx 0.3 \;{\rm g/cm^2}$,  increasing temperature ${\bar T}>6$~keV may miss the ignition threshold.  However, with ${\bar T} \approx 5.5$~keV, increasing $\rho R$ should always lead to ignition. 

The ignition boundary predicted by  Eq.~(\ref{eq:StagnationTemperature}) is sensitive to the analytic scalings of various terms.  While most researchers employ similar expressions for $Q_\alpha$ and  $Q_{rad}$, a variety of refinements have been considered for thermal conduction $Q_e$, including modifications due to dense plasma effects \cite{Hurricane_2019B} and magnetic fields \cite{Sadler_2022}.    The precise form will influence the threshold along the high-temperature boundary where thermal conduction dominates over radiation losses.  

Experimental measurements from N210207 strongly imply that the previous ignition threshold ($q_{fus}>q_{rad} + q_e$) was achieved \cite{Zylstra_2022}, yet the capsule failed to ignite.  One possible explanation is the neglect of $P dV/dt$ losses after stagnation.   However, as shown in Fig.~\ref{fig:RadiusEnergy} for N210207, there is an interval ($\sim 60$~ps) near stagnation where $P dV/dt$ is small.  During this interval, mass ablation is a fundamental limitation that must be overcome, and our results suggests that Eq.~(\ref{eq:StagnationTemperature}) represents the actual ignition boundary for spherical hot spots near stagnation.  This conjecture is consistent with N210207 \cite{Zylstra_2022} which falls short of the new threshold, while with minor changes N210808 \cite{Hurricane_2022} achieved higher $\rho R$ and crossed the threshold.   While the experiments are complicated by other factors (fill tube, low-mode perturbations, instabilities), these close neighbors feature similar differences in experimental ($\sim 8\times$) and simulated ($\sim 13\times$) yields, reflecting the importance of 1D physics.  We anticipate that low-mode perturbations of the hot spot will produce a modest increase in $dM/dt$ relative to 1D simulations.  Other multi-dimensional effects may further increase $dM/dt$ beyond the ablation driven rate, such as micro-jetting from the dense fuel into the hot spot arising from defects in the outer ablator~\cite{Haines_2022}.  

\begin{acknowledgments}
We gratefully acknowledge D. Clark, A. Kritcher, and C. Weber for providing the data needed to simulate the shots.
\end{acknowledgments}

 \appendix
\section{Mass Ablation Rate}
 \label{MassAppendix}
For a hot spot near ignition conditions (${\bar T} \sim 5$~keV, $R\sim40\;\mu {\rm m}$), the conduction term in  Eq.~(\ref{eq:EnergyConservation}) redistributes energy on fast time scales $\tau \sim R^2/{\mathcal D} \lesssim 100$~ps, where ${\cal D}$ is the diffusion coefficient defined below.  While radiation also influences the temperature profile on these fast time scales, it is more difficult to treat analytically.  To understand the temperature $T(r,t)$ and mass density $\rho(r,t)$ profiles within isobaric hot spots, we consider the conservation equation for the internal energy with only thermal conduction
\begin{equation}
\rho \left ( \frac{\partial \varepsilon}{\partial t} + {\bf U} \cdot \nabla \varepsilon \right ) + p \nabla \cdot {\bf U} =  q_e \;,
\end{equation}
where $\varepsilon = C_V T$ is the specific internal energy, ${\bf U}$ is the fluid velocity, $p$ is the pressure and $C_V$ is the specific heat at constant volume.  On the fast time scales of interest ($t<\tau$ ), we neglect the flow terms and assume the isobaric pressure $P\equiv \eta \rho T$ is nearly constant, where $\eta$ is the ideal gas constant.   Eliminating density in favor of temperature, we obtain the simplified heat flow equation 
 \begin{equation}
 \frac{C_V}{\eta} {P \over T} {\partial T \over \partial t} = \nabla \cdot \kappa \nabla T\;.
\label{eq:heatconductioneqn} 
\end{equation}
We assume thermal conduction is of the form $\kappa \equiv k_0 T^\beta$ where $k_0$, $\beta$ are constants.  In previous work \cite{Springer_2019}, researchers employed this same basic equation but with the specific heat at constant pressure $C_P$.  We note that several classic textbook references \cite{Landau,Zeldovich}  on thermal conduction in subsonic fluids also have $C_P$ on the left hand side.   However, the textbooks are actually considering problems where the absolute temperature differences are small, so that the density can be taken as constant.   These assumptions are clearly not applicable to ICF hot spots, which feature large spatial variations in both the density and temperature.  Furthermore, we will demonstrate that the mass ablation rate implied by Eq.~(\ref{eq:heatconductioneqn}) is consistent with previous theoretical work \cite{Atzeni_1984, Atzeni_2004,Hurricane_2016,Hurricane_2019,Hurricane_2019B} in the limit where the dense fuel is cold (see below) and with the simulation results of the present paper.

Our goal is to relate the mass ablation rate into the hot spot in terms of the outward heat loss.  To proceed, we assume spherical symmetry and express the hot spot mass as
\begin{equation}
M(t) \equiv 4 \pi \int_0^{R(t)} \rho(r,t) r^2 dr = \frac{4 \pi P}{\eta} \int_0^{R(t)}  \frac{r^2}{T(r,t)} dr \; ,
\end{equation}   
and the ablation rate as
\begin{equation}
\frac{dM}{dt} = - \frac{4 \pi P}{\eta} \int_0^{R(t)} \frac{1}{T^2} \frac{\partial T}{\partial t} r^2 dr + 4 \pi \rho(R,t) R^2 \frac{dR}{dt} \;.
\label{dMdt1}
\end{equation}
The first term arises from the fluid velocity $U_r$ at the edge of the hot spot while the second term is due to the outward propagation of the heat front into the dense fuel (with temperature $T_\ice$).  In the limit $T_\ice \rightarrow 0$, then $dR/dt \rightarrow 0$ and Eq.~(\ref{eq:heatconductioneqn}) has a separable solution~\cite{Springer_2019}
\begin{equation}
T(r,t) = T_o(t) [ 1 - ({r / R_o} )^2]^{{1 /( 1 + \beta)}}\;,
\label{eq:Tprofile}
\end{equation}
where $T_o(t) = T_{\emptyset} \left( 1 + 6 {\cal D} t/R_o^2 \right)^{-1/(1+\beta)}$ is the central temperature, ${\cal D} \equiv \eta k_0 T_{\emptyset}^{\beta+1}/(C_V P)$, $T_{\emptyset}$ is the initial central temperature, and $R_o$ is the fixed radius where the temperature vanishes.  In this separable limit, the mass ablation rate in Eq.~(\ref{dMdt1}) is proportional to the heat flux at any radius $r<R_o$ within the hot spot
\begin{equation}
\bar{T} C_V \frac{dM}{dt} = q_e \equiv   -4 \pi R^2  \left( \kappa \frac{\partial T}{\partial r} \right)_{r=R}
\label{dMdt3}
\end{equation}
where $\bar{T}  =  PV/(\eta M)$ is the mass averaged temperature and $V=4\pi R^3/3$.  Since Eq.~(\ref{dMdt3}) is valid at any radius $r<R_o$,  one may define the hot spot radius $R(t) <R _o$ to encompass some fraction of the fusion reactivity.   However, while Eq.~(\ref{dMdt3}) is used widely \cite{Atzeni_1984,Atzeni_2004,Hurricane_2016,Hurricane_2019,Hurricane_2021}, this result is only valid for $T_\ice=0$.  Notice that $\partial T/ \partial r$ in Eq.~(\ref{eq:Tprofile}) is singular as $r \rightarrow R_o$ and $\rho \rightarrow \infty$, which prevents the heat front from propagating.    

Although Eq.~(\ref{eq:heatconductioneqn}) is not strictly separable for finite edge temperature, we can estimate the $dR/dt$ contribution by matching an approximate solution for the inner region to a boundary layer near the ice.  Here, the radius $R(t)$ is defined by the matching temperature $T_m \equiv T(R,t) $ between these regions, and the subscript $m$ is used to denote all quantities at this location.   There is freedom to choose $T_m$, provided that $T_o \gg T_m \gg T_\ice$.   Assuming the reactivity scales as $Y_n \propto T^{n-2}$, one may roughly estimate $T_m/T_o \approx  X^{1/(n-2)}$ to encompass a specified fraction $X \equiv Y_n(R)/Y_n(0)$, where $n \approx 3.5$ at $T \sim 5$~keV.     For accuracy, in this paper we employ the actual fusion reactivity $Y_n(r)$ to define $T_m$ and the corresponding hot spot radius $R(t)$. 
   
To proceed with the boundary layer analysis, we define $F(r,t) \equiv (T/T_\emptyset)^{\beta + 1}$  and re-write Eq.~({\ref{eq:heatconductioneqn}) as 
\begin{equation}
\frac{\partial F}{\partial t} = {\cal D} F \frac{1}{r^2} \frac{\partial}{\partial r} r^2 \frac{\partial F}{\partial r} \approx {\cal D} F {\partial^2 F \over \partial r^2} \;,
\label{Fdimensional}
\end{equation}
where the approximate form (far right) ignores the geometric terms within the boundary layer $r \gtrsim R(t)$ near the ice.  Within the hot spot $r<R(t)$, the approximate solution to Eq.~(\ref{Fdimensional}) is
\begin{equation}
\label{newF}
F(r,t) \approx F_o(t) \left [1 - (1-\xi_m)  \left( \frac{r}{R(t)} \right)^2 \right]\;,
\end{equation}
 where $F_o \equiv (T_o/T_\emptyset)^{\beta + 1}$  and $\xi_m \equiv (T_m(t)/T_o(t))^{\beta+1}$  is a constant defining $R(t)$.   Neglecting the geometric terms within the boundary layer,  Eq.~(\ref{Fdimensional}) is invariant to a translation Lie group~\cite{Dresner_1999} and has traveling wave solutions of the form $F(r - u t)$ where $u=dR/dt$ is the speed of the ablation front into the ice.  Substituting into \refeq{Fdimensional}, we obtain $F'' = -(u/{\cal D}) F' / F$. 
Integrating and applying boundary conditions at $x \rightarrow \infty$, we get $F' = -(u/{\cal D})\ln(F/F_\ice)$, 
where $F_\ice \equiv (T_\ice/T_\emptyset)^{\beta + 1} $ for the asymptotic fuel temperature $T_\ice$ deep in the 
ice layer. Matching to $dF/dr$ of the inner solution as $r \rightarrow R(t)$, we solve for the ablation speed
\begin{equation}
\label{dRdt1}
\frac{dR}{dt} \equiv u = \frac{2 {\cal D} (1-\xi_m) F_o(t)} {R(t) \ln ( F_m(t)/F_{ice} )} \;.
\end{equation}
Expressing $F_o$ in terms of the heat flux 
\begin{displaymath}
q_e \equiv   -4 \pi R^2  \left( \kappa \frac{\partial T}{\partial r} \right)_{r=R} = 8 \pi R  (1- \xi_m) \frac{k_0 T_\emptyset^{\beta +1}}{\beta+1} F_o \;,
\end{displaymath}
we can now include both terms in Eq.~(\ref{dMdt1}) for the mass ablation rate
\begin{equation}
\bar{T} C_V \frac{dM}{dt} =  \left[ 1 +  \frac{\rho_m/\bar{\rho}}{\ln (T_m(t)/T_\ice)}\right] q_e \equiv {\mathcal A} q_e \;, 
\label{eq:Adefinition}
\end{equation}
where $\bar{\rho}/\rho_m =T_m/\bar{T} = H(\xi_m) \xi_m^{1/(1+\beta)}$, $\rho_m \equiv \rho(R,t)$ is the edge density, $\bar{\rho} \equiv M/V$ and
\begin{equation}
H(\xi_m) \equiv   3 \int_0^{1}  \frac{x^2}{\left[ 1- (1-\xi_m) x^2 \right]^{1/(1+\beta)}} dx  \;.
\end{equation}
The dimensionless factor ${\mathcal A} = 1 +  (\rho_m/\bar{\rho})/\ln (T_m/T_\ice)$ represents the enhancement in the mass ablation into hot spot arising from the heat loss into the dense fuel.  In the limit $T_\ice \rightarrow 0$, this factor limits to ${\mathcal A} \rightarrow 1$ in agreement with Eq.~(\ref{dMdt3}).
 
\subsection{Numerical Verification } 
To understand the accuracy of Eq.~(\ref{eq:Adefinition}), we solved Eq.~(\ref{eq:heatconductioneqn}) numerically as an initial value problem using an explicit finite-difference method.   The initial condition was specified by Eq.~(\ref{newF}) for $r<R$ and by the boundary layer solution outside this radius.  Numerical convergence was carefully checked with respect to the radial and temporal discretization.  In the numerical solutions, we set $\beta=0.67$ consistent with the simulation profiles shown Fig.~\ref{fig:Tprofile} and also with radiation-hydrodynamic simulations performed with the HYDRA code \cite{Patel_2020}.   Note that the conduction scaling exponent is typically $\beta \approx 2 - 2.5$ in hot plasmas. The lower value of $\beta=0.67$ needed to match the simulations is likely due to the radiation transport, which is not accounted for in the above analysis.   Since the xRAGE simulations employ the same opacity and conduction models as typically employed with HYDRA,  it is not surprising that both give the same value of $\beta=0.67$ to fit the temperature profile.  In the experiments, there are a variety of factors that may impact the hot spot temperature profile, including uncertainties in the opacity and thermal conduction, and also 3D perturbations to the hot spot.   However, notice that the factor ${\mathcal A}$ does not depend explicitly on $\beta$, which implies the basic prediction may be more general (see Fig.~\ref{fig:MassTemp} for comparison with 1D radiation-hydrodynamics simulations).  Since the hot spot radius $R(t)$ is defined by a threshold on $T_m(t)/T_o(t)$, the time evolution of $T(r,t)$ can be used to numerically evaluate $M$, ${\bar T}$, $q_e$ and the factor
\begin{equation}
{\mathcal A}  \equiv   \frac{ \bar{T} C_V}{q_e} \frac{dM}{dt}  \;.  
\end{equation}
The top panel of Fig.~\ref{fig:TheoryNumerical} compares time evolution of ${\mathcal A}(t)$ obtained from the numerical solution (red) against Eq.~(\ref{eq:Adefinition}) (blue dashed) for a hot spot radius $R(t)$ defined by $T_m/T_o=0.3$ and for the range of $T_\ice$ values shown.   The numerical solution features a short period in which the initial condition relaxes followed by slow evolution of ${\mathcal A}(t)$ which is in reasonable agreement with the analytic estimate.  The bottom panel of Fig.~\ref{fig:TheoryNumerical} compares the scaling of ${\mathcal A}(t)$ with $T_\ice$ for three different threshold values defining the hot spot radius $T_m/T_o=0.2, 0.3, 0.4$.  These results demonstrate that Eq.~(\ref{eq:Adefinition}) is valid regardless of whether the hot spot radius is defined by a constant $T_m/T_o$ or by a reactivity threshold $X$ for which $T_m/T_o$ varies in time.

 \begin{figure}
 \includegraphics[width=\linewidth]{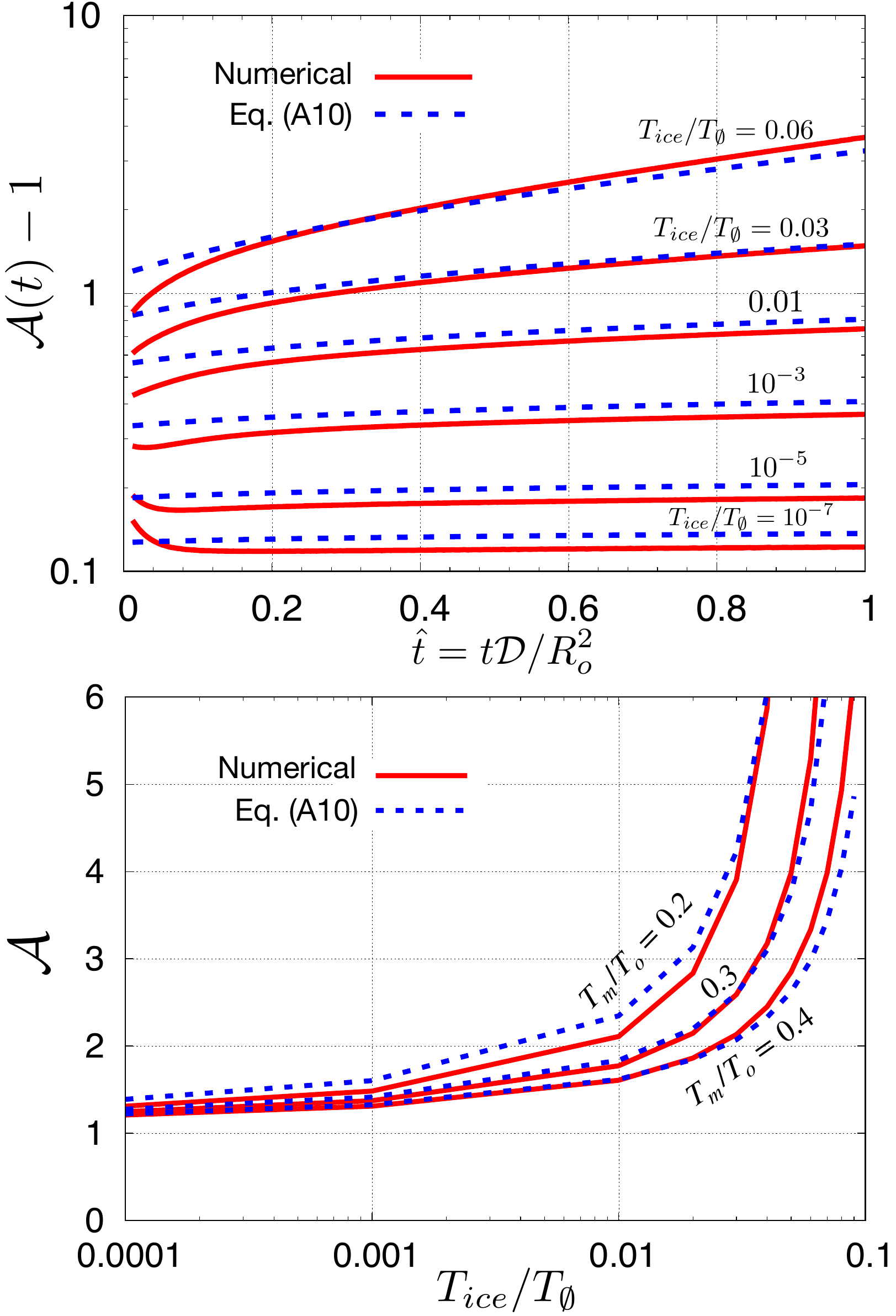}
 \caption{Top panel compares the evolution of ${\cal A}(t)$ from Eq.~(\ref{eq:Adefinition}) against ${\mathcal A}$ inferred from the numerical solution of Eq.~(\ref{eq:heatconductioneqn}) using $T_m/T_o=0.3$ to define $R(t)$.  Bottom panel compares the scaling of ${\mathcal A}$ as a function of $T_\ice$ between the numerical solution (red) and Eq.~(\ref{eq:Adefinition}).  This comparison is shown at $\hat{t} \equiv t {\cal D}/R_o^2 = 1$ for different values of $\xi_m \equiv (T_m(t)/T_o(t))^{1+\beta}$ defining the hot spot radius. The conduction scaling exponent is $\beta = 0.67$, consistent with the observed simulation profiles shown in Fig.~\ref{fig:Tprofile}.  \label{fig:TheoryNumerical}}
 \end{figure}  

 \section{Ignition Threshold}  
 \label{IgnitionAppendix}
 The static ignition threshold is obtained from (\ref{eq:StagnationTemperature}) by requiring $d{\bar T}/dt>0$ and introducing approximate scalings for the various terms.  Here, we follow Ref.~\cite{Atzeni_2004} and express the fusion power as $Q_\alpha = c_\alpha g_\alpha {\bar \rho}   \langle \sigma v \rangle$ where $c_\alpha \approx 8.1 \times10^{40}\; {\rm erg/g^2}$, $\langle \sigma v \rangle$ depends only on temperature \cite{Bosch_1992}, and $g_\alpha \approx 1.2$ is obtained from averaging $\langle \sigma v \rangle$ over the isobaric profile.   We employ the newly developed parameterization \cite{Zylstra_2019} of the $\alpha$ particle range and deposition fraction $f_\alpha$.   Based upon two modern theories \cite{Maynard_1985, Brown_2005} of charged particle stopping power, the new parameterizations \cite{Zylstra_2019} are consistent with both experiments and simulations. For hot spots near ignition, the resulting $f_\alpha$ is slightly lower than older results \cite{Krokhin_1973}. While in Fig.~\ref{fig:Ignition} we employ the fit to Ref.~\cite{Brown_2005},  the parameterization of Ref.~\cite{Maynard_1985} gives very similar results. 
  
 The bremsstrahlung radiation loss is $Q_{rad} = c_b {\bar \rho} {\bar T}^{1/2}$ where $c_b \approx 3.05\times10^{23}\;{\rm  erg\, cm^3/(g^2 \,s \, keV^{-1/2}})$, and we assume conduction losses are dominated by electrons 
\begin{equation}
Q_e  \approx \frac{\kappa_e | \nabla T_e | (4 \pi R^2)}{M} \sim \frac{3 c_e {\bar T}^{7/2}}{\ln \Lambda \, {\bar \rho} R^2} \;,
\end{equation}
Here, we have assumed $\nabla \; \sim R^{-1}$ and use Spitzer conductivity $\kappa_e  = c_e {\bar T}^{5/2}/\ln \Lambda$ where $c_e \approx 9.67\times10^{19}\;{\rm erg/(s\, cm \, keV^{7/2})}$, and $\ln \Lambda \approx 3.5$ is the Coulomb logarithm.   Substituting these scalings into (\ref{eq:StagnationTemperature}), the ignition threshold $d{\bar T}/dt>0$ can be written as
\begin{equation}
{\bar \rho} R > \left( \frac{3 c_e {\bar T}^{7/2} \ln \Lambda^{-1} (1+D)}{c_\alpha g_\alpha \langle  \sigma v \rangle \left[f_\alpha -D \left(1-f_\alpha \right) \right] - c_b {\bar T}^{1/2}} \right)^{1/2}  .
\label{eq:IgnitionThreshold}
\end{equation}
Neglecting weak dependencies within $\ln \Lambda$, the ignition threshold is a function of $\bar{\rho} R$ and ${\bar T}$ together with the conditions in the dense fuel ($T_\ice$).   The results in Fig.~\ref{fig:Ignition} are presented in terms of $\rho R \equiv \int_0^R \rho dr \approx 0.79 {\bar \rho} R$.   


\end{document}